# Venusian phosphine: a 'Wow!' signal in chemistry?

*William Bains[1,2,#,*], Janusz J. Petkowski[1,#,*], Sara Seager[1,3,4], Sukrit Ranjan[5,6,7], Clara Sousa-Silva[8], Paul B. Rimmer[9,10,11], Zhuchang Zhan[1], Jane S. Greaves[2], Anita M. S. Richards[12]*

[1]Dept. of Earth, Atmospheric, and Planetary Sciences, Massachusetts Institute of Technology, 77 Mass. Ave., Cambridge, MA, 02139, USA.

[2]School of Physics & Astronomy, Cardiff University, 4 The Parade, Cardiff CF24 3AA, UK

[3]Dept. of Physics, Massachusetts Institute of Technology, 77 Mass. Ave., Cambridge, MA, 02139, USA.

[4]Dept. of Aeronautics and Astronautics, Massachusetts Institute of Technology, 77 Mass. Ave., Cambridge, MA, 02139, USA.

[5]Northwestern University, Center for Interdisciplinary Exploration and Research in Astrophysics, Evanston, 60201, USA

[6]Northwestern University, Department of Astronomy & Astrophysics, Evanston, 60201, USA

[7]Blue Marble Space Institute of Science, Seattle, 98154, USA

[8]Harvard-Smithsonian Center for Astrophysics, Observatory Building E, 60 Garden St, Cambridge, MA 02138, USA

[9]Department of Earth Sciences, University of Cambridge, Downing Street, Cambridge CB2 3EQ, UK.

[10]Cavendish Laboratory, University of Cambridge, JJ Thomson Ave, Cambridge CB3 0HE, United Kingdom

[11]MRC Laboratory of Molecular Biology, Francis Crick Ave, Cambridge CB2 0QH, United Kingdom

[12]Jodrell Bank Centre for Astrophysics, Department of Physics and Astronomy, The University of Manchester, Alan Turing Building, Oxford Road, Manchester, M13 9PL, UK.

# These authors contributed equally to this work, and are listed alphabetically.

* Correspondence to: bains@mit.edu, jjpetkow@mit.edu.

## Abstract

The potential detection of ppb levels phosphine ($PH_3$) in the clouds of Venus through millimeter-wavelength astronomical observations is extremely surprising as $PH_3$ is an unexpected component of an oxidized environment of Venus. A thorough analysis of potential sources suggests that no known process in the consensus model of Venus' atmosphere or geology could produce $PH_3$ at anywhere near the observed abundance. Therefore, if the presence of $PH_3$ in Venus' atmosphere is confirmed, it is highly likely to be the result of a process not previously considered plausible for Venusian conditions. The source of atmospheric $PH_3$ could be unknown geo- or photochemistry, which would imply that the consensus on Venus' chemistry is significantly incomplete. An even more extreme possibility is that strictly aerial microbial biosphere produces $PH_3$. This paper summarizes the Venusian $PH_3$ discovery and the scientific debate that arose since the original candidate detection one year ago.

## Graphical Abstract

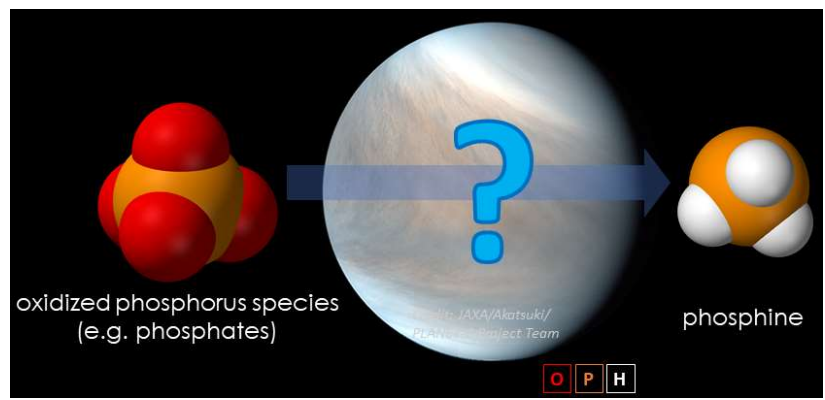

## Introduction

Venus is about the same size and mass as Earth and is sometimes called Earth's sister planet. However unlike the bulk planet composition Venus' atmospheric chemistry and surface conditions are quite different from Earth's (Smrekar et al., 2014). Carbon dioxide ($CO_2$) is a dominant gas in the atmosphere of Venus with molecular nitrogen ($N_2$) making most of the rest of volatiles (up to 5% (Peplowski et al., 2020)). The trace gas composition of the atmosphere of Venus is complex, with contributions of water vapor ($H_2O$), carbon monoxide (CO), sulfur dioxide ($SO_2$), hydrogen chloride (HCl) and many other species (Fegley, 2014). The Venusian clouds and hazes are known to have a complex vertical atmospheric profile with several distinct layers. The main cloud layer (~48 to ~70 km) is composed of droplets, which are believed to be made primarily of photochemically produced sulfuric acid (Oschlisniok et al., 2012). Haze extends from below the clouds through the cloud layer to at least 100 km and may be composed of elemental sulfur as well as sulfuric acid (Taylor et al., 2018; Titov et al., 2018). However, our understanding of the chemistry of the Venusian atmosphere and clouds is incomplete. In particular Venus atmospheric phosphorus chemistry is a neglected topic. Phosphorus species were detected in Venus' atmosphere by the VeGa-1 and VeGa-2 probes during their descent through the clouds towards the surface of the planet (Andreichikov, 1987 a; b), however their chemical identity remains unknown. Phosphorus chemistry in the atmosphere and the clouds of Venus only recently got more attention (Bains et al., 2021 a; Milojevic et al., 2021; Mogul et al., 2021 a) following the tentative detection of phosphine ($PH_3$) gas (Greaves et al., 2021 a).

## Venusian phosphine detection one year later

The recent candidate $PH_3$ detection in the Venus' cloud decks adds further questions to the already complex picture of the chemical composition of the atmosphere of Venus (Greaves et al., 2021 a).

Even one year after the candidate detection of few ppb of $PH_3$ in the atmosphere of Venus (Greaves et al., 2021 a) the discovery draws much interest and controversy. The original potential $PH_3$ detection has been based on a single-millimeter wavelength absorption line, the $PH_3$ 1–0 rotational transition at 1.123 mm wavelength, observed by two independent facilities, both James Clerk Maxwell (JCMT) and Atacama Large Millimeter Array (ALMA) telescopes, at 5 and 7 sigma confidence levels respectively. Since the initial $PH_3$ discovery was announced, several papers have questioned the detection, either on the grounds of data analysis (Snellen et al., 2020; Thompson, 2021; Villanueva et al., 2021) or an assignment of mesospheric $SO_2$ rather than cloud-level $PH_3$ (Akins et al., 2021; Lincowski et al., 2021). In addition, several groups have used IR observations to provide strong upper limits (in the low ppb-sub ppb range) on the abundance of $PH_3$ above the clouds (Encrenaz et al., 2020; Trompet et al., 2020); however these observations do not test whether $PH_3$ is present in or below the clouds.

The reanalysis of the data by several methods, including a non-subjective method of characterizing the datasets that departs from the usual polynomial fitting procedures (Greaves et al., 2021 b; c) support the initial detection. The authors of the original discovery provided a response to the critiques, both on data processing and data interpretation, demonstrating that $PH_3$-identification was not a post-hoc rationalization of a feature found after complex data processing (Greaves et al., 2021 b; c; d).. As an internal control of applied methods an expected absorption line of deuterated water (DHO), in right abundance, has been detected with the same procedures.

Several authors suggested that the signal from JCMT telescope can be attributed to mesospheric $SO_2$ instead of $PH_3$ (Lincowski et al., 2021; Villanueva et al., 2021). It is however unlikely that the contested absorption line comes from the $SO_2$; simultaneous observation of other $SO_2$ absorptions show that the potential $SO_2$ line-contamination is less than <10% of the observed signal (Greaves et al., 2021 d).

An independent re-analysis of the legacy data collected by the Pioneer Venus Neutral Gas Mass Spectrometer (LNMS) (Mogul et al., 2021 a) for the altitude of 51.3 km shows evidence of $PH_3$ in the clouds of Venus, via detection of $PH_3$ fragmentation ions: $^+P$, $^+PH_3$, $^+PH_2$, and $^+PH_2D$. The assignment of $PH_3$ rests primarily on the detection of the phosphorus ion $^+P$. $^+P$ is most strongly associated with $PH_3$, for two key reasons. First, $PH_3$ is the only P-containing molecule that fits the data, and is in gas form at Venus' 51.3 km altitude. One might argue there could be a very tiny amount of e.g. phosphoric acid vapor that could have fragmented into $^+P$, but the corroborating fragmentation ions were not detected and likely would have been. Second, $^+P$ does not overlap with any other neutral gas mass fragment expected from the Venus atmosphere, giving $^+P$ a unique and robust detection. The reanalysis yields a $PH_3$ abundance in the mid-to-high ppb range.

Thus, there is strongly suggestive evidence from two independent methods that there is phosphine in the cloud decks of Venus. However, the debate on the presence of $PH_3$ in the clouds of Venus continues and will likely only be resolved by future *in situ* measurements of $PH_3$ gas in the Venus atmosphere.

## Phosphine on Venus cannot be explained by canonical processes

The detailed analysis of photochemical and other endergonic chemistry that could produce $PH_3$ under Venus conditions carried out by Bains et al 2021 (Bains et al., 2021 a) confirms that none of the modelled kinetic pathways can explain the levels of phosphine observed, falling short by many orders of magnitude, even using the most conservative assessments available (see Table 1). Bains et al 2021 calculations are exclusive to gas phase photochemistry, solid photochemistry was not explored as no significant UV penetrates to the surface of the planet (see below for discussion of UV photocatalytic processes in relation to mineral dust at the cloud level).

Similarly to kinetic analyses summarized above the thermodynamic analyses carried out by Bains et al 2021 show that none of the known possible routes for production of $PH_3$ on Venus can explain the presence of ~1 ppb phosphine. All fall short, often by many orders of magnitude (Table 1; Figure 1). The thermodynamics of known reactions between chemical species in the atmosphere and on the surface of Venus is too energetically costly and cannot be responsible for the spontaneous formation of $PH_3$ in the amounts detected (some processes can generate $PH_3$, but only at much lower abundances). The formation of $PH_3$ in the subsurface is also not favored. Oxygen fugacity ($fO_2$) of the crustal and mantle rocks is many orders of magnitude too high to support reduction of mineral phosphates to $PH_3$. Hydrolysis of phosphide minerals, both from crustal and mantle rocks, as well as delivered by meteorites, also cannot provide sufficient amounts of $PH_3$ to explain the observations (Bains et al., 2021 a).

**Table 1.** Summary of main potential canonical $PH_3$ production pathways on Venus (see (Bains et al., 2021 a) for the full detailed analysis).

| Potential Canonical $PH_3$ Production Pathway on Venus | Barriers for $PH_3$ Production Pathway |
|---|---|
| Photochemical production by photochemically-generated reactive species | The required forward reaction rates are too low by 5 orders of magnitude. |
| Equilibrium thermodynamics of chemical reactions between chemical species in the atmosphere and on the surface | Chemical reactions in Venusian environment are on average 100 kJ/mol too energetically costly (10 - 400 kJ/mol) to proceed spontaneously assuming 1 ppb $PH_3$. |
| Equilibrium thermodynamics of chemical reactions in the subsurface | Oxygen fugacity ($fO_2$) of plausible crust and mantle rocks is 8 - 15 orders of magnitude too high to support reduction of phosphate. |
| Phosphides from crustal and mantle minerals | Near-surface chemistry is highly unlikely to be compatible with phosphide rocks. Deep mantle phosphides will only be exposed on the surface by planetary-scale mantle over-turn events. |
| Meteorites as a source of phosphine | The estimated maximal yearly meteoritic delivery of $PH_3$ is ~8 orders of magnitude too low to explain detected amounts. |
| Production by lightning | Limited frequency of lightning and low abundance of both atmospheric phosphorus species and reducing gases. Less than ppt of $PH_3$ is produced. $PH_3$ production is ~7 orders of magnitude too low to explain detected amounts. |
| Large-scale comet/asteroid impact | Radar mapping of the surface of Venus that shows no evidence of a recent, sufficiently large, impact. The ablation of large phosphide-rich impactors can only result in transient $PH_3$ production if at all. |
| Other endergonic processes: solar X-rays and solar wind protons | Solar X-rays and solar wind protons are absorbed at high altitudes, and so could not penetrate to the clouds where phosphorus species might be found and where $PH_3$ is detected (see main text for the potential formation of $PH_3$ in the mesosphere). |
| Other endergonic processes: tribochemical processes | Maximal efficiency of formation of $PH_3$ by large tribochemical processes is 2 orders of magnitude too low to explain the detected amounts and requires efficient plate tectonics and abundance of $H_2O$ in the crust that Venus does not appear to have in sufficient amounts. |
| Known exotic chemistry as a source of $PH_3$ (e.g. formation of $PH_3$ from elemental phosphorus or production of $PH_3$ with reducing agents more powerful than $H_2$) | Such scenarios just replace the implausibility of making $PH_3$ with another implausible set of conditions which could then produce $PH_3$. |

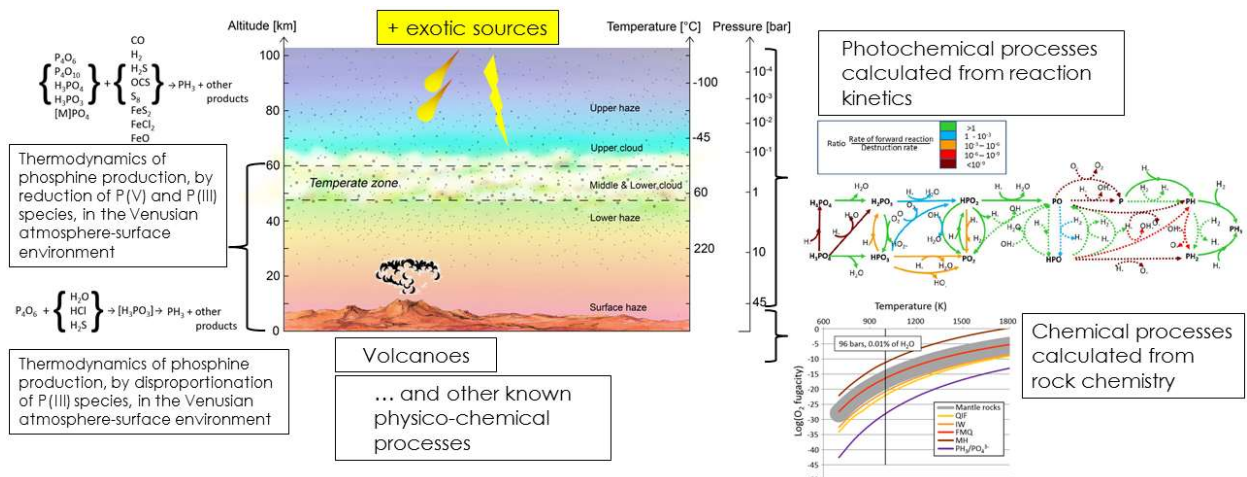

**Figure 1.** An overview of canonical processes that could in principle produce PH$_3$ on Venus. None of the examined processes produce sufficient amounts of PH$_3$ to explain the observed 1 ppb abundance (see (Bains et al., 2021 a) for the detailed analysis).

Several studies proposed potential mechanisms for non-biological PH$_3$ production on Venus. Truong and Lunine 2021 suggest that deep plume mantle volcanism could be responsible of Venusian PH$_3$ (Truong et al., 2021), but the deep plume mantle volcanism as a source of PH$_3$ is unlikely (see summarized in (Bains et al., 2021 a) and expanded upon in (Bains et al 2022 in prep.)). Omran et al 2021 suggest ablation of phosphide-rich large impactors as transient source of PH$_3$ in Venus' atmosphere (Omran et al., 2021). In this scenario the ablated phosphide rich minerals would hydrolyze in sulfuric acid environment of the clouds transiently releasing PH$_3$ in the process (Omran et al., 2021). If phosphides were delivered to the clouds from space, it is likely they would be oxidized by concentrated sulfuric acid (which is a strong oxidizing agent (Bains et al., 2021 a)), not hydrolyzed to PH$_3$. If Venus' clouds droplets have much lower concentration of H$_2$SO$_4$ than expected, as recently suggested by (Mogul et al., 2021 b), then in principle such transient PH$_3$ release could happen. However an ablating iron/nickel bolide (the most likely carrier for Fe$_3$P minerals), with a surface temperature of >1500 °C (Lovering et al., 1960) would not react with cloud droplets but with their thermal decomposition products SO$_2$, H$_2$O and O$_2$ (Atomics, 1985), a reaction that would not generate PH$_3$. Alternative scenario of PH$_3$ production on Venus proposed by Omran et al 2021 involves disproportionation of P$_4$O$_6$ (Omran et al., 2021), which have been shown to be thermodynamically implausible by (Bains et al., 2021 a).

## Non-canonical processes producing phosphine on Venus

If the PH$_3$ detection is correct and if no conventional chemical processes can produce phosphine on Venus then there has to be a not yet considered process or set of processes that could be responsible for PH$_3$ formation.

One of the possibilities is that chemical species or physico-chemical processes exist in the crust, or in the atmosphere of Venus, that have not been considered to be possible on rocky planets like Venus.

A specific example of such a process would be exotic chemistry in the sulfuric acid cloud droplets. The chemistry of phosphorus species in concentrated sulfuric acid is rarely studied. Phosphorus oxyacids act as *bases* in concentrated sulfuric acid, forming protonated forms quite unlike those that are stable in water (Sheldrick, 1966, 1967) . Similarly phosphine is protonated in concentrated sulfuric acid; even though its pKa as measured in water (Weston Jr et al., 1954) suggests that it is too weak a base to be protonated in an acid with a Hammett acidity of -11.5 (Liler, 1971), NMR studies clearly show it is present as phosphonium ($PH_4^+$) ions (Figure 2).

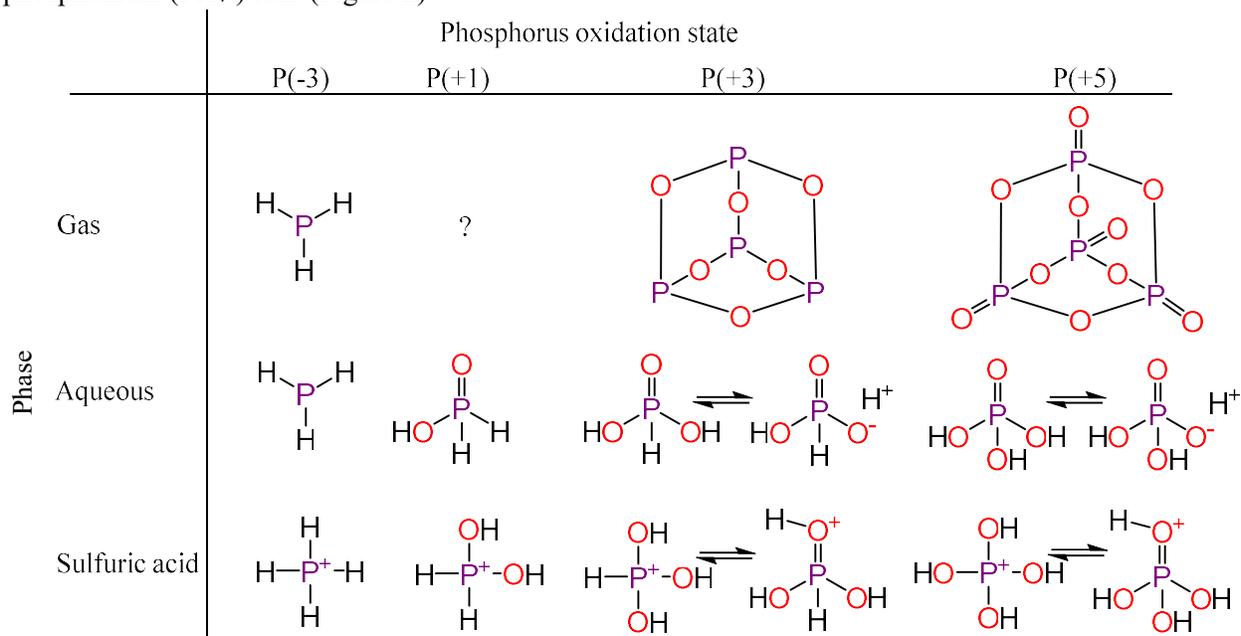

**Figure 2.** Structures of phosphorus compounds in gas, water and concentrated sulfuric acid phases. Structure data for gas and water phases from (Greenwood et al., 2012). Structures of phosphorus compounds for 98% sulfuric acid phase from (Sheldrick, 1966, 1967). The structure of compounds in concentrated sulfuric acid can be very different from those in aqueous solution. Note that the P(+1) forms also have resonant or ionization equivalents (not shown).

Likewise, the photochemistry of phosphorus species in sulfuric acid droplets is completely unknown, so hypothetically $PH_3$ could be produced photochemically in the sulfuric acid droplets of the cloud layer. $PH_3$ is rapidly oxidized by sulfuric acid to phosphoric acid at temperatures > 0 ºC. Even if a photochemical process did produce $PH_3$ in sulfuric acid droplets, it seems unlikely that phosphine would escape oxidation back to phosphoric acid. Nevertheless, one of the non-canonical processes that could potentially lead to the formation of reduced phosphorus species, including $PH_3$, could involve photocatalytic reduction of phosphorus oxides, like $P_4O_6$. The analogous mineral-dependent, UV-photocatalyzed reduction reactions have been studied for oxidized nitrogen and carbon species.

For example, $NH_3$ can be photochemically produced on iron-doped $TiO_2$-containing sands (Kasting, 1982; Schrauzer et al., 1983). When exposed to sunlight or UV, $N_2$ and $H_2O$, $TiO_2$-containing sands can reduce $N_2$ to $NH_3$ and trace $N_2H_4$. On Earth such abiotic $N_2$ photocatalytic-fixation is only observed in arid and semiarid regions, like deserts, but not in areas with abundant water (Schrauzer et al., 1983). Similarly, photocatalytic synthesis of $CH_4$ from $CO_2$ and $H_2$ over acidic minerals has been studied (Civiš et al., 2016; Knížek et al., 2020). Several minerals facilitate photocatalytic formation of $CH_4$ from $CO_2$, including $Al_2O_3$, ilmenite ($FeTiO_3$) (Knížek et al., 2020).

It is unknown if photocatalytic reduction processes analogous to those of $N_2$ and $CO_2$ can also proceed for phosphorus oxides like $P_4O_6$ (Kaiserová, 2021), but the extremely dry and acidic environment of the Venusian clouds makes such reactions in principle possible. Photocatalytic reduction of $P_4O_6$ on various mineral dust surfaces can be tested experimentally under simulated Venusian cloud conditions (temperature, pressure and UV radiation regimes etc.). The efficiency of the transport of the mineral dust to the clouds and the stability of the minerals, in the Venusian surface and cloud environment, required to facilitate $P_4O_6$ reduction should also be assessed.

The presence of unexpected minerals, or expected minerals at unexpected locations, on Venus that could act as powerful reducing agents is a testable hypothesis that could be the subject of future remote or *in situ* observation.

There also remains a possibility that $PH_3$ is present in the mesosphere, above the clouds (Greaves et al., 2021 d; Lincowski et al., 2021). This is unlikely, as photochemical destruction of $PH_3$ above the cloud is very fast, with half-times of seconds (Bains et al., 2021 a), and no $PH_3$ was detected above the clouds by IR observation (Encrenaz et al., 2020; Trompet et al., 2020). However, if $PH_3$ is indeed present that high up in the atmosphere then other photochemical processes could conceivably generate $PH_3$ through processes not dependent on mineral catalysts. $PH_3$ as a product of solar X-rays and solar wind protons could remain as a possibility as those phenomena carry substantial energy to potentially drive $PH_3$ synthesis, although the availability of phosphorus species that could be substrates for $PH_3$ formation in the mesospheric regions is unknown. We note however that if the $PH_3$ is made in the mesosphere above the clouds, then its presence would have to be reconciled with the possible detection of $PH_3$ by Pioneer Venus in the cloud decks (Mogul et al., 2021 a).

## The challenges for the biological production of $PH_3$ on Venus

Phosphine is a biosignature gas on Earth (Bains et al., 2019 a; b; Sousa-Silva et al., 2020), and has been proposed as a biosignature gas on exoplanets (Sousa-Silva et al., 2020). The announcement of the potential presence of $PH_3$ on Venus resulted in the community showing a small but significant increase in the notion that life in the clouds of Venus is a possibility (Bains et al., 2021 b). Nevertheless the community still considers Venus to be the least likely abode of life ranking far behind Europa, Enceladus, Mars and even behind Titan (Bains et al., 2021 b). This skepticism results from the fact that Venusian environment has many challenges for life as we know it (Seager et al., 2021).

The challenges for life in the Venus clouds are unique and cannot be directly compared to the challenges that life faces in Earth's extreme environments (Seager et al., 2021). Even if the Venusian clouds are possibly much more clement, and less acidic than previously thought (Bains et al., 2021 c; Mogul et al., 2021 b), one of the most challenging aspects of Venusian cloud environment is lack of hydrogen. Biochemicals are hydrogen-rich molecules, which makes the synthesis of any biochemical in a low hydrogen environment thermodynamically challenging. More specifically, we have very little idea as to why life would invest rare hydrogen to produce phosphine in large amounts which it then 'throws away' by releasing it into the atmosphere (Benner, 2021). The model describing a plausible biological production of $PH_3$ presented in (Bains et al., 2021 a) is solely a model of the thermodynamics of the process, not a model of the biochemistry or a modelled metabolic pathway, thus the actual biosynthetic mechanism or the evolutionary fitness that a hypothetical $PH_3$-producing organism gains by producing $PH_3$ in an severely H-depleted environment is unknown.

## Conclusions

One year after the original announcement, the tentative discovery of $PH_3$ in the clouds of Venus continues to bring much interest and controversy. The $PH_3$ 1–0 rotational transition at 1.123 mm wavelength is the only $PH_3$ transition feasible to observe from the surface of the Earth. In the future, higher-frequency transitions may be accessible for observation from space telescopes. The $PH_3$ observation requires confirmation, e.g. by the detection of additional phosphine spectral features, and the debate on $PH_3$ on Venus will likely only be resolved by future *in situ* measurements. If confirmed the detection of $PH_3$ in the Venus' atmosphere requires an explanation. The source of $PH_3$ is not known, but could be unknown geo- or photochemistry, which would imply that the current understanding of the Venusian planetary processes is significantly incomplete, or even more unexpectedly that PH$_3$ is a product of a strictly aerial microbial biosphere.

# Acknowledgements

We thank the Heising-Simons Foundation and the Change Happens Foundation for funding. S.R. acknowledges the funding from the Simons Foundation (495062). C.S.-S. acknowledges the 51 Pegasi b Fellowship and the Heising- Simons Foundation. P.B.R. acknowledges funding from the Simons Foundation (SCOL awards 599634).

# Author Disclosure Statement

No competing financial interests exist.